# Strain-modulated Slater-Mott crossover of pseudospin-half square-lattice in $(SrIrO_3)_1/(SrTiO_3)_1$ superlattices

Junyi Yang[1,*], Lin Hao[1,*], Derek Meyers[2], Tamene Dasa[3], Liubin Xu[3], Lukas Horak[4], Padraic Shafer[5], Elke Arenholz[5,6], Gilberto Fabbris[7], Yongseong Choi[7], Daniel Haskel[7], Jenia Karapetrova[7], Jong-Woo Kim[7], Philip J. Ryan[7], Haixuan Xu[3], Cristian D. Batista[1], Mark P. M. Dean[2], Jian Liu[1,*]

1. Department of Physics and Astronomy, University of Tennessee, Knoxville, Tennessee 37996, USA
2. Department of Condensed Matter Physics and Materials Science, Brookhaven National Laboratory, Upton, New York 11973 USA
3. Department of Material Science and Engineering, University of Tennessee, Knoxville, Tennessee 37996, USA
4. Department of Condensed Matter Physics, Charles University, Ke Karlovu 5, 121 16 Prague, Czech Republic
5. Advanced Light Source, Lawrence Berkeley National Laboratory, Berkeley, CA 94720, USA
6. Department of Materials Science & Engineering, University of California, Berkeley, CA 94720, USA
7. Advanced Photon Source, Argonne National Laboratory, Argonne, IL, 60439 USA

*To whom all correspondence should be addressed.

jyang43@vols.utk.edu; lhao3@utk.edu; jianliu@utk.edu

Abstract

We report on the epitaxial strain-driven electronic and antiferromagnetic modulations of a pseudospin-half square lattice realized in superlattices of $(SrIrO_3)_1/(SrTiO_3)_1$. With increasing compressive strain, we find the low-temperature insulating behavior to be strongly suppressed with a corresponding systematic reduction of both the Néel temperature and the staggered moment. However, despite such a suppression, the system remains weakly insulating above the Néel transition. The emergence of metallicity is observed under large compressive strain but only at temperatures far above the Néel transition. These behaviors are characteristics of the Slater-Mott crossover regime, providing a unique experimental model system of the spin-half Hubbard Hamiltonian with a tunable intermediate coupling strength.

Electron-electron interaction holds the key to numerous emergent phenomena of modern condensed matter physics, such as superconductivity, insulator-to-metal transition, quantum magnetism, colossal magnetoresistance, stripe order, and spin liquid phases [1-5]. The idea that a sufficiently large Coulomb repulsion triggers collective localization of the electrons and opens a correlated charge gap in an otherwise metallic system, has been widely used to account for intriguing insulating states in a huge variety of quantum materials. Such a correlated gap opening is often accompanied with the emergence of magnetism [3,6]. A prominent example is the Mott insulating parent compound of high-$T_c$ cuprates [7,8], where the localized electrons interact with each other through superexchange interactions and form antiferromagnetic (AFM) order below the Néel temperature $T_N$. The key physics of such many-body behavior is well captured by the single-band two-dimensional (2D) Hubbard Hamiltonian on a square lattice [9-13]. While this picture essentially maps the half-filled Hubbard Hamiltonian to the Heisenberg Hamiltonian in the limit of strong Coulomb repulsion [7,8], it is known treating the Hubbard Hamiltonian in the weak Coulomb repulsion limit also stabilizes an insulating ground state simply driven by the AFM order, i.e., the so-called Slater insulator [14]. Despite yielding the same ground state, these two perturbative approaches at the two opposite limits predict drastically different behaviors in the paramagnetic state: a Slater insulator is metallic above $T_N$, whereas a Mott insulator remains insulating. This distinction highlights the fact that solving the 2D Hubbard Hamiltonian is highly challenging, despite its simple form, especially in the regime of intermediate interaction and finite temperatures, where there is no small control parameter and unlike the Slater and Mott approaches. It is thus crucial to obtain and drive real 2D systems across this regime in experiments.

The recent advances in the field of 5$d$ iridates have led to new opportunities with the so-called $J_{eff}$ = 1/2 electrons, which can also be described by an effective spin-half

Hubbard Hamiltonian [15-17]. The local $J_{eff}$ = 1/2 Kramer doublet is stabilized by strong spin-orbit coupling (SOC) under an octahedral crystal field and is half-filled under the $Ir^{4+}$ $5d^5$ low-spin configuration [15,18-23]. The resulting $J_{eff}$ = 1/2 band on a square lattice indeed exhibits an AFM insulating ground state in $Sr_2IrO_4$ [19,20,24,25] and $Ba_2IrO_4$ [26-28]. Both systems share similar structural motifs to the cuprate parent compound $La_2CuO_4$ [7]. But the larger spatial extension of the $5d$ orbitals also has reduced correlation from that of $3d$ orbitals, implying that iridates may fall into the intermediate-coupling regime [29,30]. Indeed, resistivity and optical conductivity measurements have suggested a much smaller charge gap in the iridates [31,32], opening the door to driving and examining a 2D half-filled single-band system across the intermediate interaction regime. The insulating behavior of $Sr_2IrO_4$ turns out to be fairly robust against high pressure applied up to 55 GPa [33,34]. $Ba_2IrO_4$, on the other hand, was found to become metallic around 13 GPa [35]. However, the crucial response of the AFM order remains unclear in both cases, although the weak ferromagnetism of $Sr_2IrO_4$ due to spin canting disappears around 20 GPa [34].

In this work, we present a systematic investigation of the stability of the AFM order and the electronic modulation of the $J_{eff}$ = 1/2 square lattice in $(SrIrO_3)_1/(SrTiO_3)_1$ superlattice (SL) by varying epitaxial strain. As shown in Fig. 1(a), this SL structure is effectively an artificial crystal of $Sr_2IrTiO_6$ [32,36,37], where the square lattice of $IrO_6$ octahedra is separated by a $SrTiO_3$ monolayer, mimicking the quasi-2D $SrIrO_3$ layers in the Ruddlesden-Popper structures [38-42]. When grown on a $SrTiO_3$ substrate, this SL exhibits a $J_{eff}$ = 1/2 AFM insulating state [43-45], similar to $Sr_2IrO_4$ and $Ba_2IrO_4$ [19,20,26,35]. Both theoretical and experimental studies have found that the low-energy electronic structure, the intralayer magnetic structure and interactions of the SL are analogous with $Sr_2IrO_4$ and can be described by a half-filled effective spin-half Hubbard Hamiltonian [36,37,44,46]. By increasing the compressive epitaxial strain, here we find the SL shows a weakened insulating behavior and becomes a bad metal in the high-temperature paramagnetic phase. Meanwhile, all the studied SLs are found to have an AFM ground state at low temperatures with systematically decreasing ordering temperatures and ordered moments. The high-temperature metallicity and the low-temperature AFM ordering are bridged by a weakly insulating regime at intermediate temperatures. Through polarization dependent x-ray absorption measurements and first-principles calculations, we verify that the strain induced modulation is driven by a reduced effective correlation due to the enhanced in-plane Ir-O hybridization. The extracted temperature-strain dependence unravels the emergent behaviors when modulating a prototypical 2D Hubbard system across the Slater-Mott crossover regime.

The SLs were fabricated by pulsed laser deposition with *in-situ* reflection high energy electron diffraction. More details of the growth and characterizations can be found in Refs. [43,47]. We tune the epitaxial strain by growing the SL on three different substrates [Fig. 1(b)]: SrTiO$_3$ (001) (STO, $a_{pc}$ = 3.905 Å), (LaAlO$_3$)$_{0.3}$(Sr$_2$TaAlO$_6$)$_{0.7}$ (001) (LSAT, $a_{pc}$ = 3.868 Å) and NdGaO$_3$ (001) (NGO, $a_{pc}$ = 3.863 Å). During the growth, all the SLs were kept in the same stacking sequence with a thickness of 30 supercells by *in-situ* monitoring the deposition process. X-ray diffraction (XRD) measurements were performed on a Panalytical X'Pert MRD diffractometer to verify the crystalline quality as well as the epitaxial relationship. Synchrotron XRD measurements and x-ray resonant magnetic scattering (XRMS) experiments were performed at beamlines 33BM and 6IDB, respectively, at the Advanced Photon Source of the Argonne National Laboratory. A crystal analyzer was adopted during the XRMS measurement to improve the magnetic signal-to-noise ratio. X-ray absorption spectroscopy (XAS) experiments at the Ir L-edge were performed at beamline 4IDD at the Advanced Photon Source of the Argonne National Laboratory. These measurements confirm that the picture of the half-filled pseudospin-half state is valid for our samples [47]. Linear polarization-dependent XAS experiments at the O K-edge were performed at beamline 4.0.2, at the Advanced Light Source of the Lawrence Berkeley National Laboratory. First-principle density functional theory calculations were performed using the projector augmented wave technique as implemented in Vienna ab-initio Simulations Package [55]. Due to the large SOC of Ir, we have carried out non-collinear spin-dependent calculations to determine the electronic properties. More details about the calculation can be found in Supplemental Material [47].

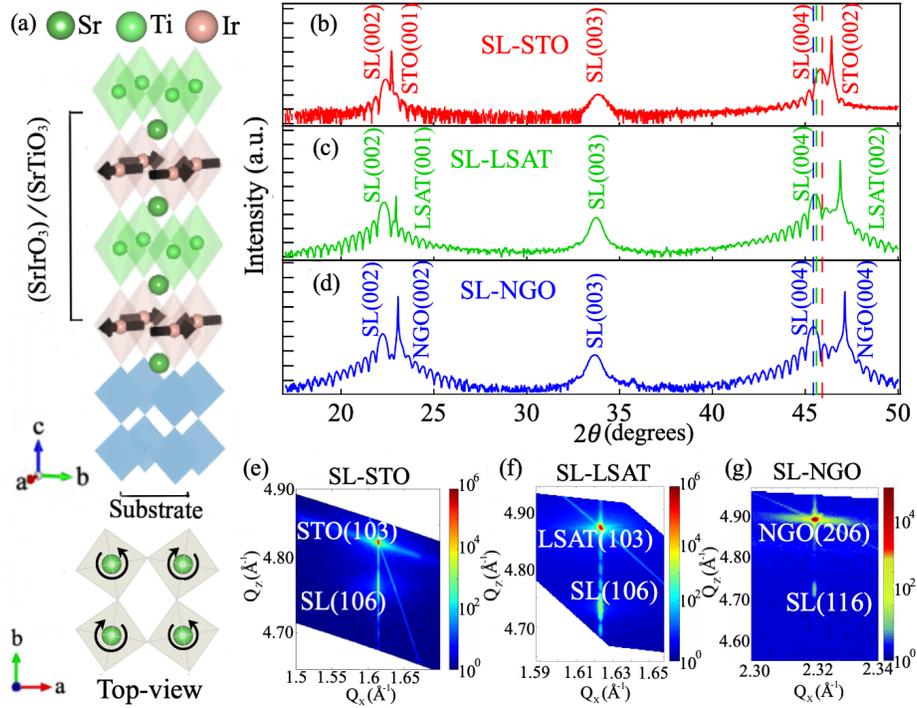

FIG. 1. (a) A schematic of the $(SrIrO_3)_1/(SrTiO_3)_1$ superlattice grown on a substrate with compressive epitaxial strain. Because of the in-plane compression, the lattice structure is elongated along the out-of-plane [001] direction. The black arrow denotes the spin. Top view of the square lattice of $IrO_6$ octahedra. Rotation around the $c$-axis causes a $\sqrt{2}\times\sqrt{2}$ cell expansion of the square lattice. (b) - (d) XRD patterns along the (0 0 L) direction for SLs grown on STO, LSAT and NGO, respectively. The supercell $a\times a\times 2c$ ($a$ and $c$ are pseudo-cubic in-plane and out-of-plane lattice parameters, respectively) is used for the notation. The blue, green and red dashed lines represent the (0 0 4) film peak position of SL-NGO, SL-LSAT and SL-STO, respectively. (e) - (g) Reciprocal space mappings around the film (106) or (116) reflection of SL-STO, SL-LSAT and SL-NGO. The same in-plane Q-vectors of the SLs and the corresponding substrates demonstrates that all the SLs are fully strained within the experimental resolution.

Figure 1(b) to (d) show the representative XRD *θ-2θ* scans for the SLs grown on the three substrates, respectively. Clear (0 0 *even*) reflections with pronounced Kiessig fringes can be seen on all the SLs, indicating high epitaxial quality and sharp film-substrate interface. In addition, we observed clear (0 0 *odd*) peaks that come from the alternating SIO-STO growth mode of the SLs, confirming the realization of the as-designed stacking pattern along the *c*-axis. Since the applied strain is expected to be increasingly compressive from STO, to LSAT, and to NGO, it is important to verify the strain state of the SL. We performed reciprocal space mapping (RSM) to measure the in-plane lattice parameters. As shown in Fig. 1(e) to (g), the Bragg peaks of all SLs are fully aligned with that of the underlying substrates along the in-plane direction with no observable asymmetric intensity distribution, demonstrating the fully strained state. Therefore, the in-plane lattice parameter of the SLs decreases monotonically going from STO to LSAT, and NGO substrates. Correspondingly, the (0 0 *L*) peak positions of the SLs are systematically shifted to lower angles, indicative of an expansion along the *c*-axis. The extracted pseudocubic *c*-axis lattice parameter of the SLs increases from 3.954 Å on STO (SL-STO) to 3.980 Å on LSAT (SL-LSAT) and 3.988 Å on NGO (SL-NGO). An important structural distortion in the physics of the 2D $J_{eff}$ = 1/2 electrons is the octahedral rotation/tilting [22,56,57]. We performed synchrotron-based XRD to measure the corresponding half-order structural peaks [58]. The results show that SL-LSAT and SL-NGO only have octahedral rotation with respect to the *c*-axis while octahedral tilting with respect to the *a*- or *b*-axis is not observable [47]. This 2D $IrO_6$ octahedral structure leads to a $\sqrt{2} \times \sqrt{2}$ cell expansion within the *ab*-plane (Fig. 1(b)), similar to that in $Sr_2IrO_4$ [59]. SL-STO also has a significant octahedral rotation, but a small octahedral tilting is observable as well [47], consistent with the previous reports [46,60]. The disappearance of the octahedral tilting in SL-LSAT and SL-NGO indicates that the planar oxygen ions, while displaced, remain in the same plane as the Ir sites and the Ir-O-Ti bond is straightened by compressive strain [61].

With increasing compressive strain, the SLs show a systematic suppression of resistivity and insulating behavior. Figure 2(a) compares their temperature dependent resistivity from 300 K to 10 K. As can be seen, the resistivity of the SL-STO increases by about two orders of magnitude upon cooling to the base temperature, whereas the SL-LSAT increases only about one order. When further increasing the compressive strain, the resistivity of SL-NGO was further reduced. More importantly, its temperature dependence displays a metallic behavior at high temperatures. As shown in the inset of Fig. 2(a), the resistivity monotonically decreases with temperature in the high-temperature region until reaching a minimum around 180 K, below which the resistivity slowly increases for only three times down to 10K. This observation suggests an emerging "bad metallicity" [62] of the $J_{\text{eff}}$ = 1/2 electrons in the SL under large compressive strain, which is in contrast to the robust insulating behavior in $Sr_2IrO_4$ under high pressure [33,34] or compressive strain [63-65].

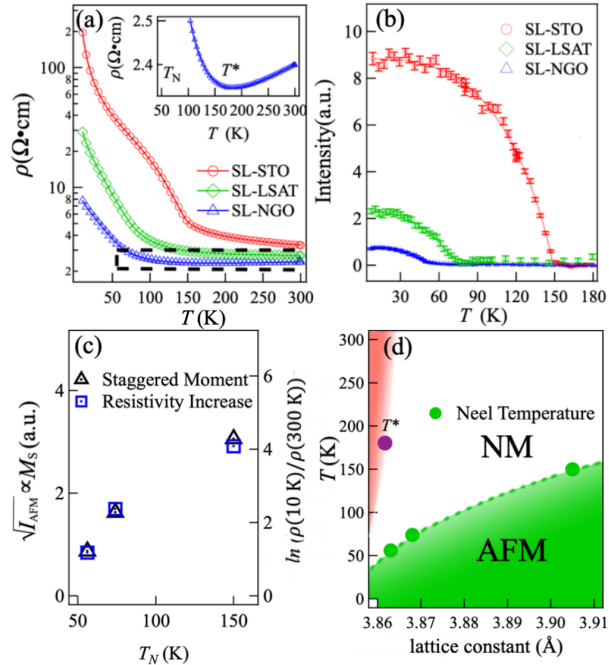

FIG. 2. (a) Temperature dependent resistivity of SL-STO (red circles), SL-LSAT (green diamonds) and SL-

NGO (blue triangles). The inset shows an enlarged view of the highlighted portion (dashed box) of the resistivity curve for the SL-NGO. (b) Temperature dependence of the normalized (0.5 0.5 2) magnetic peak intensity of SL-STO, SL-LSAT and SL-NGO. The superlattice cell $a \times a \times 2c$ was used to define the reciprocal space. (c) The evolution of staggered moment and resistivity increases are plotted against the Néel temperature $T_N$. (d) A summary diagram of the phase evolution of the SL with respect to temperature and in-plane lattice constant. The green region denotes antiferromagnetic (AFM) insulating state with the green dashed line being the phase boundary, while the white and red regions represent non-magnetic (NM) insulating and metallic states, respectively.

Upon the suppression of the insulating state and the emergence of metallicity, the fate of the AFM order is crucial for revealing the underlying mechanism. While AFM order is often probed by neutron scattering in bulk crystals, it is highly challenging for ultrathin films due to the small sample volume. Instead, we performed XRMS measurements at the Ir $L_3$-edge on the SLs to directly monitor the AFM Bragg peak. We observed (0.5, 0.5, $L = even$) magnetic reflections at 7 K in all three SLs (see Supplemental Material [47]), demonstrating the persistence of the AFM ground state. The AFM structure of the SLs is determined to be $C$-type, where the Ir moments order antiferromagnetically within the square lattice and the adjacent Ir layers order ferromagnetically (Fig. 1(a)). Nevertheless, there is a significant and systematic strain-driven modulation of the AFM order. Figure 2(b) compares the temperature dependence of the integrated intensity of the (0.5 0.5 2) AFM Bragg peak for the SLs. The magnetic peak of SL-STO disappears above 150 K, which is consistent with the previous reports [44,66]. As the compressive strain increases, the magnetic peak for SL-LSAT and SL-NGO vanishes at 75 K and 55 K, respectively, demonstrating a systematic decrease of $T_N$. In addition, the integrated intensity of the AFM Bragg peak ($I_{AFM}$) at 7 K decreases by about 75% from SL-STO to SL-LSAT and by about another 75% from SL-LSAT to SL-NGO. Since the AFM Bragg peak is proportional to the staggered magnetization square ($M_s^2$), this measurement allows quantifying the strain dependence of the AFM order parameter, which drops by half from SL-STO to SL-LSAT and by another half from SL-LSAT to SL-NGO.

Figure 2(c) summarizes the evolution of the electronic and magnetic properties of the SLs by comparing the AFM order parameter ($\sqrt{I_{AFM}}$) and the logarithm of the resistivity increase, which characterizes the strength of the insulating behavior, as functions of the Néel temperature. In particular, the fact that $T_N$ is suppressed in accordance with the weakening of the insulating behavior demonstrates the dominant role of charge fluctuations in controlling the thermal stability of the quasi-2D AFM order. This is consistent with the observed amplitude reduction of the ground state staggered magnetization, which also signifies enhanced charge fluctuations due to delocalization. The fact that $M_s$ was reduced simultaneously with $T_N$ can be understood in the weak coupling limit of the 2D Hubbard model, where both the mean-field ordering temperature and the staggered magnetization are proportional to the ground state charge gap induced by AFM pairing. This is essentially the picture of the Slater insulating state [14]. This picture is however clearly inconsistent with the experimental observations above $T_N$. Specifically, a metallic state is necessary in the Slater picture in the absence of magnetic order, whereas all three SLs show insulating behavior above $T_N$, as summarized in Fig.2(d). Such an observation indicates a charge gap already exists before the AFM order sets in, reminiscent of the Mott insulating state in the strong coupling limit.

The blend of the behaviors characteristic of the weak and strong coupling limits clearly suggests that the system should be instead considered in the Slater-Mott crossover regime or the intermediate-coupling regime, which is the most challenging one for solving the 2D Hubbard model, especially at finite temperatures [10]. While the AFM insulating ground states of the two limits can be continuously connected [10,67], one of their key differences is the size of the magnetic moment [68]. The observed variation of the base-temperature staggered magnetization of the SLs shows a strain-driven modulation of the pseudospin-half square lattice within the crossover regime. At temperatures above $T_N$, our results indicate that the Mott gap remains finite but is relatively small, especially under a large compressive strain. The local moment is thus expected to be small and fluctuates strongly with temperature due to significant thermal excitation of the charge carriers. When temperature is comparable with the charge gap, the moment will be completely annihilated with the emergence of metallicity. This corresponds to temperatures that are usually unpractically high but could be reached in the Slater-Mott crossover regime, such as the thermal evolution of SL-NGO from the AFM ground state to the weakly insulating paramagnetic state and eventually to the weakly metallic state (Fig. 2(d)). The results of our study reveal the unique character of the crossover regime unexpected in both the Mott and Slater pictures.

To shed additional light on the strain-induced variation in the underlying electronic structure, we performed linearly polarized XAS experiments at the O K-edge, which probes the unoccupied states projected onto the O 2$p$-orbitals (Figs. 3(a)-(c)). Figure 3(d) displays the polarization dependent spectra near the absorption edge for all the SLs. The pre-edge (527-529 eV) feature is characteristic of the $J_{eff}$ = 1/2 state hybridized with the O 2$p$ orbitals, whereas the higher-energy (529-534 eV) peak represent the Ir $e_g$ band and the Ti $t_{2g}$ band [63]. We observed a clear difference of the pre-edge intensities between the in-plane and out-of-plane polarization channels, implying an anisotropic hybridization of the Ir $J_{eff}$ = 1/2 state and the O 2$p$-orbitals. In the out-of-plane channel, the x-ray probes the hybridization of the 2$p_z$ orbitals of the four planar oxygen sites with the Ir $d_{yz}$ and $d_{xz}$ orbitals (Fig. 3(a)). On the other hand, the in-plane channel is sensitive to the hybridization of the Ir $d_{xy}$ orbital with the two planar oxygen 2$p_y$ orbitals (Fig. 3(b)) and the Ir $d_{yz}$ orbital with the two apical oxygen 2$p_y$ orbitals (Fig. 3(c)). Therefore, the larger spectral intensity in the out-of-plane channel from Fig. 3(d) indicates a stronger hybridization between the Ir ions with the planar oxygens in all the SLs. Figure 3(e) illustrates the difference spectra between the out-of-plane and in-plane channels, i.e., x-ray linear dichroism (XLD). A systematic enhancement of the XLD signal can be seen from the SL-STO to SL-NGO. This observation indicates that the overall hybridization with the planar oxygen increases with increasing compressive strain, which could lead to reduction of the effective correlation of the 2D $J_{eff} = 1/2$ band and the observed modulation within the Slater-Mott crossover regime. For comparison, we performed first-principles calculations on the whole series of SL [47]. To assess the native influence of the structural strain on the electronic hybridization, we set $U$ to zero in all the calculations. Figure 3(f) and (g) presented the density of states (DOS) of the planar oxygen as well as the apical oxygen within ±0.4 eV from the Fermi level, where the $J_{eff} = 1/2$ band is located. One can see that the overall DOS of planar oxygen ions is indeed increased from SL-STO to SL-LSAT and SL-NGO, whereas the overall DOS of apical oxygen ions has been systematically reduced with increasing compressive strain, as expected from the elongated lattice along the $c$-axis. We further extracted the $p_z$ and $p_y$ components of the DOS of both oxygen ions and took a difference following the similar rule in XLD measurement. This difference of the projected DOS is shown in Fig. 3(h), where a systematic increase with compressive strain can be seen for the unoccupied states above the Fermi level, similar to the XLD result.

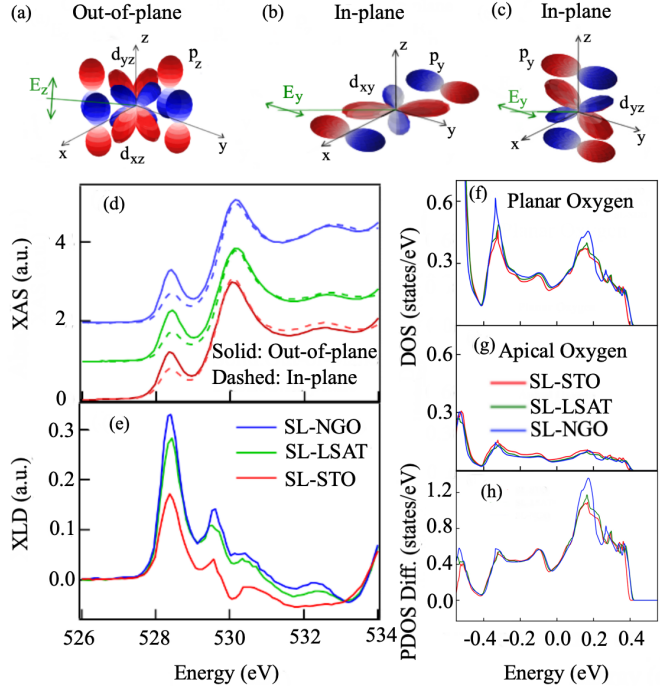

FIG. 3. Schematic diagram of linearly dependent XAS measurement at O K-edge. (a) Out-of-plane channel. (b) and (c) In-plane channel. (d) Polarization dependent O *K*-edge x-ray absorption (XAS) spectra of SL-STO (red), SL-LSAT (green) and SL-NGO (blue). The solid/dashed line denotes XAS from out-of-plane/in-plane measurement. The absorption is shifted vertically for clarity. (e) X-ray linear dichroism extracted from (d). Density of states (DOS) of the planar oxygen ions (f) and apical oxygen ions (g). (h) The projected DOS difference (PDOS Diff.) extracted from (f) and (g).

In conclusion, we investigated the epitaxial strain-induced electronic and magnetic evolution of $(SrIrO_3)_1/(SrTiO_3)_1$ SL. By increasing compressive strain, we efficiently reduced the effective correlation strength of the $J_{eff} = 1/2$ electrons. Correspondingly, the staggered magnetization and the Néel temperature are systematically and significantly suppressed though AFM structure remains robust within the range of applied strain. The insulating behavior was also strongly suppressed with the emergence of metallicity at high temperatures. When lowering the temperature, the metallic state crossovers into a weak insulating state before the AFM order kicks in. We argue this evolution of the low-temperature ground state in conjunction with the emerging high-temperature excited state is characteristic of modulation of the interaction coupling within the Slater-Mott crossover regime of a half-filled spin-half Hubbard system on a square lattice. Given the possibility of direct probe of the AFM order and epitaxial engineering, the SL represents an excellent model system for exploring the emergent phenomena in this intriguing regime.

The authors acknowledge experimental assistance from H.D. Zhou, M. Koehler. J.L acknowledges support from the Science Alliance Joint Directed Research & Development Program at the University of Tennessee and the National Science Foundation under Grant No. DMR-1848269. J.Y acknowledge funding from the State of Tennessee and Tennessee Higher Education Commission (THEC) through their support of the Center for Materials Processing. J.L., H.X., and C.D.B. acknowledge support by the Organized Research Unit Program at the University of Tennessee. M.P.M.D is supported by the U.S. Department of Energy, Office of Basic Energy Sciences, Early Career Award Program under Award No. 1047478. L.H. acknowledge the support by the ERDF (project CZ.02.1.01/0.0/0.0/15_003/0000485) and the Grant Agency of the Czech Republic grant (14-37427 G). Use of the Advanced Photon Source, an Office of Science User Facility operated for the US DOE, OS by Argonne National Laboratory, was supported by the U. S. DOE under contract no. DE-AC02-06CH11357. Work at ALS is supported by the Director, Office of Science, Office of Basic Energy Sciences, of the US Department of Energy under Contract number DE-AC02-05CH11231. Parts of the characterizations were conducted at the Center for Nanophase Materials Sciences, which is a DOE Office of Science User Facility.